\documentclass[letterpaper]{jpconf}
\usepackage{graphicx}
\usepackage{cite}

\begin{document}
\title{KATRIN: an experiment to determine the neutrino mass}

\author{R.G. Hamish Robertson}

\address{Center for Experimental Nuclear Physics and Astrophysics, University of Washington, Seattle, WA, 98195, USA}

\ead{rghr@u.washington.edu}

\author{KATRIN Collaboration}

\begin{abstract}
KATRIN is a very large scale tritium-beta-decay experiment to determine the mass of the neutrino.  It is presently under construction at the Forschungszentrum Karlsruhe, and makes use of the Tritium Laboratory built there for the ITER project.  The combination of a very large retarding-potential electrostatic-magnetic spectrometer and an intense gaseous molecular tritium source makes possible a sensitivity to neutrino mass of 0.2 eV, about an order of magnitude below present laboratory limits.  The measurement is kinematic and independent of whether the neutrino is Dirac or Majorana.  The status of the project is summarized briefly in this report. 

\end{abstract}

\section{Introduction}
In beta decay the electron and neutrino share the available energy in a statistical fashion, and in a small fraction of the decays the electron will take almost all the energy unless some must be reserved for the rest mass of the neutrino.   The phase-space available to the electron near the endpoint is therefore modified by neutrino mass, a fact realized immediately by Fermi \cite{Fermi:1934sk} as he formulated the theory of beta decay.

In the intervening seven decades of experimental searches for neutrino mass, tritium has been the beta-active nucleus of choice because it has a low endpoint energy, which makes the modification caused by neutrino mass a larger fraction of the total spectrum:  
\begin{displaymath}
  ^3{\rm H}\,\rightarrow \, ^3{\rm He}^+\,+\,{\rm e}^- \,+\,\overline{\nu}_{\rm e}  \quad {\rm +\ 18580\  eV.}
\end{displaymath}
It is also a very simple atom, and atomic or molecular effects are important at the eV level.  The decay is superallowed, with a short half-life, which reduces the amount of source material needed for a given counting rate and sensitivity. 

Oscillation experiments now define a lower limit to the mass range \cite{Yao:2006px}.  The average mass of the three eigenstates must lie between 20 meV and 2300 meV, the lower limit arising in the `normal' hierarchy if the lightest mass is also negligibly small, and the upper limit being set by the Mainz experiment on the beta decay of tritium \cite{Kraus:2004zw}.   Above 200 meV the mass spectrum becomes essentially degenerate, and the resulting beta spectrum is indistinguishable from what  a single massive `electron neutrino'  would produce if mass eigenstates were also flavor eigenstates.     The dependence of the spectral shape on mass is given
by a phase space factor only. Moreover, the 
mass measured is independent of whether the
neutrino is a Majorana or a Dirac particle.   A neutrino mass in this quasi-degenerate regime that KATRIN is designed to explore would be of cosmological importance, having a major influence on the formation of large-scale structure in the universe.

\section{The KATRIN Experiment}

\begin{figure}
\begin{center}
\includegraphics[width=6in]{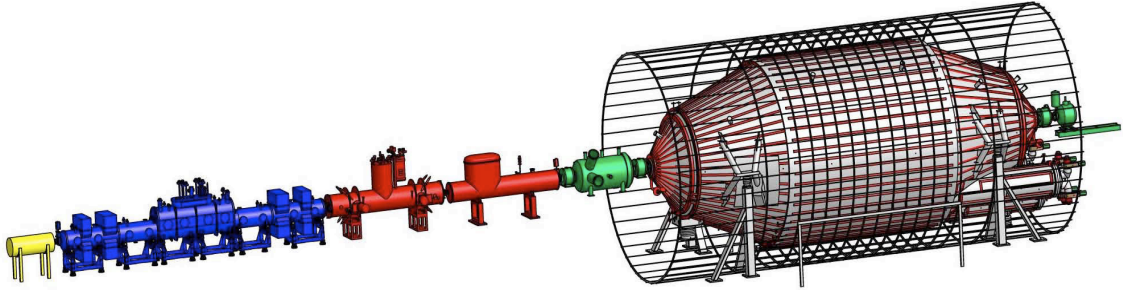}
\end{center}
\caption{\label{fig:overview}Layout of the KATRIN Experiment}
\end{figure}

The KArlsruhe TRItium Neutrino experiment \cite{Angrik:2005ep} consists of seven major subsystems (see Fig. \ref{fig:overview}), a gaseous tritium source, the tritium processing and recirculation system, a differential pumping section, a cryogenic Ar frost pumping section, a pre-spectrometer, a main spectrometer, and the detector system.  

\section{Source and Tritium Recirculation}
The gaseous molecular source consists of a tube 9 cm in diameter and 10 m in length maintained at the desired temperature (27 K) by circulation of two-phase Ne.  An axial magnetic field of 3.6 T guides electrons toward the spectrometers.  The source is being constructed by ACCEL Instruments GmbH in two stages, the first being a demonstrator to verify the stringent thermal performance specifications, including temperature regulation to $\pm30$ mK, and the second the complete system with superconducting magnets, to be delivered in early 2009.  Four turbomolecular pumps at the ends of the source tube collect T$_2$ for recirculation in an `inner loop' with Pd filters to remove contaminants, particularly $^3$He.
\section{Differential Pumping Section}
Further reduction of tritium pressure is achieved in this section, which consists of a tube with a chicane, superconducting solenoids to guide the electrons, and four 2800 l/s magnetic-bearing turbomolecular pumps to scavenge the tritium.   The subsystem is being built by Ansaldo Superconduttori for delivery in 2008.
\section{Cryogenic Pumping Section}
Any remaining tritium that escapes through the differential pumping section is trapped in Ar frost, which forms a highly efficient, large-area, chemically inert, and radiation-immune surface.  The tender for this system is about to be let.
\section{Pre-spectrometer}
There are two spectrometers in tandem in KATRIN,  both of the retarding-potential type. The pre-spectrometer operates at a cutoff potential typically 100 eV below the endpoint, preventing most electrons from reaching the main spectrometer.  Electrons can ionize residual gas molecules and create slow electrons that are indistinguishable from the signal. Reducing the electron flux into the main spectrometer is expected to improve the background near the endpoint substantially.  The pre-spectrometer was the first KATRIN subassembly to be completed, in order to permit some key concepts to be tested before other design elements were frozen.  One important discovery was the existence of a parasitic Penning-trap configuration near the ends, which made it impossible to apply high voltage and a magnetic field simultaneously without breakdown.  The addition of appropriate  electrodes suppressed this discharge completely, and the spectrometer now runs uneventfully at  35 kV and 4 T. A new feature in this type of spectrometer is a grid inside the shell to suppress the emission of low-energy electrons ejected by cosmic rays and radioactivity from the shell.  The performance of the grid  will be tested shortly, to verify the design principles of the much larger grid system for the main spectrometer.  
\section{Main Spectrometer}
The main spectrometer is a large stainless-steel vessel 10 m in diameter and 24 m in length.  The 200-tonne chamber was fabricated in Deggendorf by MAN-DWE and shipped via the river Danube, the Black Sea, the Mediterranean, the North Sea, and the Rhine to be offloaded at Leopoldshafen near the FZK laboratory.  It was placed in its building November 29,  2006.  Thermal insulation and heating tubes were installed, and 6 turbopumps. The tank was baked out to 350 C from July 16, 2007 to July 25.  The base pressure 2 months later was $0.9\times10^{-9}$ mbar and the outgassing rate (predominantly H$_2$) $1.2\times10^{-12}$ mbar l s$^{-1}$ cm$^{-2}$.  When the interior grid structure and 1000 m of non-evaporable getter strips are installed later, the predicted base pressure will be $3\times10^{-11}$ mbar.  
\section{Detector}
Electrons surmounting the potential barriers in the spectrometers enter a high axial magnetic field region and are detected in a monolithic 148-pixel Si PIN diode array 10 cm in diameter.  Two superconducting solenoids can produce up to 6 T, defining the electron beam diameter and the maximum pitch angle accepted from the source.  The magnets are being made by Cryomagnetics, Inc. and the detector by Canberra. Between the solenoids is a region in which various calibration devices -- gamma sources and a photoemissive electron gun -- can be inserted.  Selection of materials, shielding,  and an active veto are being adopted to keep non-beam-associated backgrounds below 1 mHz.  
\section{Summary}
The KATRIN experiment is scheduled for initial data-taking in 2010.  Subsystems are in an advanced state of construction and commissioning.  Funding for the project is being provided by the Helmholtz Gemeinschaft, the Bundesministerium f\"{u}r Bildung und Forschung, and the US Department of Energy. 

\section*{References}
\bibliographystyle{iopart-num}
\bibliography{rghr_KATRIN}{}

\end{document}